\documentclass[10pt,journal,compsoc]{IEEEtran}

\usepackage{epigraph}
\usepackage{physics}

\def \ee {\begin{equation}}
\def \eee {\end{equation}}
\def \eqe {\begin{eqnarray}}
\def \eqee {\end{eqnarray}}
\def \sat {{\rm sat}}

\usepackage{graphicx}

  \usepackage{cite}
\ifCLASSINFOpdf
\else
\fi
\usepackage{amsmath}
\usepackage{amsfonts}
\begin{document}
%

\title{Spiking control systems
\IEEEcompsocitemizethanks{\IEEEcompsocthanksitem Department
of  Engineering, University of Cambridge, Cambridge, UK. E-mail: rs771@cam.ac.uk}
}
%
%
%
%
\author{Rodolphe~Sepulchre,~\IEEEmembership{Fellow,~IEEE.}}
\IEEEtitleabstractindextext{%
\begin{abstract}
Spikes and rhythms organize control and communication in the animal world, in contrast to the bits and clocks  of digital technology. 
As continuous-time signals that can be counted, spikes have a mixed nature. This paper reviews ongoing efforts  to develop a control theory of spiking systems. The central
thesis is that the mixed nature of spiking results from a mixed feedback principle, and that a control theory of mixed feedback can be grounded in the operator theoretic concept of maximal monotonicity. As a nonlinear generalization of passivity, maximal monotonicity acknowledges at once the physics of electrical circuits, the algorithmic tractability
of convex optimization, and the feedback control theory of incremental passivity. We discuss the relevance of a theory of spiking control systems in the emerging age of event-based technology.
\end{abstract}
\begin{IEEEkeywords}
Feedback control, neuroscience, event-based control, passive and active electrical circuits.
\end{IEEEkeywords}}
\maketitle

\IEEEdisplaynontitleabstractindextext

%
\IEEEpeerreviewmaketitle

\ifCLASSOPTIONcompsoc
\IEEEraisesectionheading{\section{Introduction}\label{sec:introduction}}
\else

\section{Introduction}

\label{sec:introduction}
\fi

\setlength\epigraphwidth{.5\textwidth}
\epigraph{{\it Biological information-processing systems operate on completely
different principles from those with which most engineers
are familiar. For many problems, particularly those in which the
input data are ill-conditioned and the computation can be specified
in a relative manner, biological solutions are many orders of
magnitude more effective than those we have been able to implement
using digital methods. This advantage can be attributed principally
to the use of elementary physical phenomena as computational
primitives, and to the representation of information by the
relative values of analog signals, rather than by the absolute values
of digital signals. This approach requires adaptive techniques to
mitigate the effects of component differences. This kind of adaptation
leads naturally to systems that learn about their environment.
Large-scale adaptive analog systems are more robust to component
degradation and failure than are more conventional
systems, and they use far less power. For this reason, adaptive analog
technology can be expected to utilize the full potential of waferscale
silicon fabrication.}}{Carver Mead, IEEE Proceedings, 1990 \cite{mead1990}}

\IEEEPARstart{I}{n} the digital age, control systems have been  divided into distinct categories: physical systems and automata \cite{wiki:control}. Physical control systems model the continuous change over time of electrical, mechanical, or  other analog signals determined by physical laws.  Feedback control is used to regulate their function in the presence of uncertainty and to make them adaptive to a changing environment. Automata model the discrete sequence of switches of a finite state machine ruled by the logical laws of an algorithm. Physical systems and automata obey distinct modeling, analysis, and design principles. They are taught in distinct departments and researched in distinct journals and  conferences.  How to interconnect physical systems and automata has become a key hurdle of control theory, and this challenge is addressed using hybrid models and hybrid theories, that concatenate the distinct languages of physics and logic \cite{goebel2012}.

\begin{figure}[h]
\centering
\includegraphics[width=3.5in]{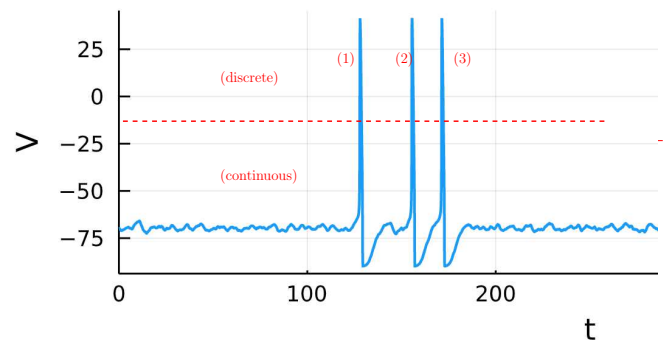}
\caption{The mixed nature of spiking signals. Spiking signals are electrical analog signals. Yet spikes can be enumerated.}
\label{fig_spiking}
\end{figure}

The divide between physical systems and automata does not apply to the animal world. The nature of control and communication in plants and animals is pulsatile, or {\it spiky}.  In nervous systems, spikes  are electrical signals that change continuously,  yet spikes can be counted. Temporal events mix the analog nature of physical systems and the digital nature of automata. We aim at developing a control theory of spiking systems that acknowledge their mixed nature: spiking control systems are {\it both} physical systems {\it and} automata, rather than a  hybrid concatenation of physical systems and automata. Spiking control systems obey physical laws and can be continuously regulated, but they aim at the reliability of digital communication.

While creating the field of neuromorphic engineering, Carver Mead envisioned a post-digital age in which the bits and clocks of digital computers would  be replaced by the spiky events and rhythms of analog electronic circuits.  He envisioned this new technology as a response to the inefficiency of digital machines in comparison to the animal world \cite{mead2020}. Thirty years later, this prediction is coming true. The end of "Moore's law" is now happening and digital technology is becoming unsustainable \cite{strubell2019}.  Carver Mead's vision and the current development of neuromorphic engineering is a response to that challenge.  We aim at  developing a control theory of spiking systems, that will apply both to the natural world of biophysical neuronal circuits and to the post-digital technology of event-based systems. 

The paper is organized as follows. Section \ref{sec:spiking} grounds spiking systems in the nonlinear circuit theory of negative conductance devices. The mathematical concept of maximal monotonicity provides a system theoretic definition of circuits with {\it positive} or {\it negative} conductance. Section \ref{sec:mixedfeedback} defines spiking control systems as {\it mixed} feedback systems, grounding the mixed nature of spiking in the mixed nature of the feedback controller. Section \ref{sec:analysis} shows that such mixed feedback systems can be analysed as difference of monotone operators. The take-home message is that the conventional analysis of physical control systems can be regarded as a theory of negative feedback systems, leading to analysis of monotone operators, whereas spiking control systems are mixed feedback systems. The paper exploits the key property that the {\it inverse} of a mixed feedback system is the difference of two maximally monotone operators. A parallel is drawn with difference-of-convex (DC) programming, which leverages the theory of convex optimization to minimize nonconvex functions modelled as a difference of convex functions. Section \ref{sec:design} addresses the design of spiking control systems, stressing the apparent conflict between the variability of analog control systems and the digital reliability of automata. The paper ends with a short discussion about the potential of spiking control theory for the design of event-based physical systems.

\section{Physical models of spiking circuits}
\label{sec:spiking}

\subsection{Electrical circuits, positivity, and passivity}

Circuit theory \cite{chua1987} is the physical modeling language of spiking systems. Neuronal activity is recorded by means of electrical signals. 
Hodgkin and Huxley used  an electrical circuit  to model the biophysics of neuronal excitability\cite{Hodgkin1952}. 
Neuromorphic electronic circuits have a circuit representation \cite{mead1989}. 

Electrical circuits model relationships between currents ($I$) and voltages ($V$). The {\it behavior} \cite{willems2007}  of the circuit is the set of current and voltage trajectories (
that is, temporal signals) $I(t)$ and $V(t)$ that obey its physical laws at every time instant $t$. The basic circuit element is a two terminal one-port,
see Figure \ref{fig:spiking_circuits}.
The port determines how the element can be interconnected and exchange energy with its environment according to Kirchhoff laws. The product $VI$ is the 
electrical power and its integral over time 
measures the total energy supplied to the device. This quantity is always positive for passive elements, meaning that the
element dissipates energy over time. For an ideal resistor $V= R I$, the supplied energy is the heat dissipated in the element. 

From an operator-theoretic perspective, passivity is closely related to a property of positivity. 
The circuit element is then regarded as an input-output operator in a Hilbert space $H$ equipped  
with an inner product $\bra{\cdot}\ket{\cdot}: X \times X \rightarrow \mathbb{R}$.  
An operator on $X$ is said to be   positive, if for
any  input-output pairs $(V,I)$, the inner product $\bra{I}\ket{V}$ is positive. 
If we choose $X=L_2(-\infty,\infty)$ the space of square-integrable functions and the standard
inner product $\bra{I}\ket{V}=\int_{-\infty}^{+\infty} I(t)^T V(t) dt$, then the positivity of the operator is closely
related to passivity. In fact, the two properties are equivalent for causal operators \cite{Desoer1975}. 

Passivity is central to the linear theory of RLC circuits.  All passive systems admit the physical realization 
of a port interconnection of  basic passive elements such as  the linear resistor $V=RI$, the linear capacitor
$C \frac{dV}{dt}=I$ and the linear inductor $L \frac{dI}{dt}=V$ \cite{Bott1949}.

Circuit theory is a parent of control theory. The system theoretic generalization of passivity is dissipativity, a concept introduced
by Willems to model physical systems defined as 
interconnections of elements that exchange energy with their environment, such as mechanical or themodynamical
systems \cite{willems1972,willems1972a,vanderschaft2014}. The linear theory of dissipativity is algorithmic, because analysis and design questions  have the formulation
of convex optimization problems, more specifically via the solution of Linear Matrix Inequalities \cite{willems1972a,boyd1994}. 
Dissipativity is central to control theory because it bridges physical modeling and the algorithmic treatment of analysis and design questions.

Spiking physical circuits cannot be passive. They require {\it active} elements, such as batteries,
to allow for self-sustained oscillations or multiple steady-state equilibria.  The mixed nature of spiking arises from
the physical interconnection of passive {\it and} active elements. A key challenge of a control theory of spiking is
to leverage the algorithmic tractability of linear passivity theory to nonlinear systems that do not just dissipate energy but 
include the regenerative elements required to spike and oscillate.

\subsection{Nonlinear resistors, monotonicity, and incremental passivity}

The concept of (maximal) monotonicity is central to this paper. Monotonicity is the {\it incremental} form of positivity: the positivity
property is not between $I$ and $V$, but between increments $\Delta I$ and $\Delta V$. 
Monotonicity was introduced by Minty \cite{Minty1960,Minty1962} to formalise and generalize the concept of physical
resistor to devices that can be nonlinear and dynamical. 
An operator on $X$ is said to be monotone, or incrementally positive, if for
any  input-output pairs $(V_1,I_1)$ and $(V_2,I_2)$, the positivity condition
$$ \bra{ I_1-I_2}\ket{V_1-V_2} \; \ge \; 0 $$
holds.

The link between monotonicity and incremental passivity is the same as  between
positivity and passivity. The two concepts are equivalent for causal operators \cite{Desoer1975,chaffey2021}. 
Given any  trajectory $(I_2,V_2)$, the
incremental supply of energy  $\bra{ I_1-I_2}\ket{V_1-V_2}$  required to generate the perturbed trajectory  $(I_1,V_1))$ 
is always positive. It reduces to passivity if one imposes the  trajectory $(I_2,V_2)$ to be the zero trajectory.

The reader will note the difference between passivity (or positivity)
and incremental passivity (or monotonicity).  A nonlinear resistor is passive 
if its graph is in the first and
third quadrant. It is incrementally passive if it is the graph of a monotone function.
The two properties only coincide for linear operators.

Following the influential paper of Rockafellar \cite{Rockafellar1976}, 
monotone operators have become a cornerstone of
convex analysis and optimization theory \cite{Rockafellar1997}. This is because the subgradient 
of a convex function always define a maximally monotone operator. As a consequence,
minimizing a convex function is equivalent to finding a zero of a maximally monotone operator.
Splitting algorithms to solve that question have known  a surge of interest in the last decade, due to their applicability
to large-scale and nonsmooth problems \cite{Ryu2016,Ryu2020}.

Operator monotonicity is a fundamental bridge between the physics of
electrical circuits, the algorithmic tractability of convex optimization, and
the feedback control theory of incremental passivity. The interested reader
is referred to \cite{chaffey2021,chaffey2021a} for more details.
In short,  the concept of monotonicity  is instrumental in generalizing 
the methodology of circuit theory  from linear to nonlinear circuits.

The reader will note that the operator theoretic concept of monotonicity in this paper differs from 
the dynamical systems  concept of monotonicity introduced by Hirsch, see e.g. \cite{Hirsch2006,Angeli2003}.
Both monotonicity concepts refer to an incremental form of positivity, but the positivity of a dynamical system
is the preservation of an order property by the flow.

\subsection{Negative resistance circuits}

\begin{figure}[!t]
\centering
\includegraphics[width=1.5in]{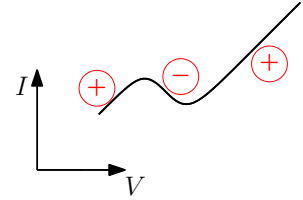}
\caption{The $IV$ curve of a negative resistance circuit, characterized by a restricted
voltage range over which the slope (i.e. resistance, or conductance) is negative.}
\label{fig:neg_res}
\end{figure}

Central to spiking is the negative resistance device shown in Figure \ref{fig:neg_res}. The device
is said to have a negative resistance (or conductance) when a positive increment of current $\Delta I$
can correspond to a negative increment of voltage $\Delta V$. 

By definition, a negative resistance device cannot be monotone. Instead, it is is  the {\it difference} of two monotone operators.  
Note that  a negative resistance device is passive
if  the graph of  its $IV$ curve is entirely in the first and third quadrant.  

Every system in this paper is defined by positive and negative interconnections of monotone operators.
 Monotone operators generalize the concept of
nonlinear resistor whereas differences of monotone operators generalize the concept
of negative resistance device. The input-output relationship of a monotone operator
can be nonlinear and dynamical. The current and voltage variables can be scalar variables,
but they can also be vector-valued. Input-output monotone operators can also be spatiotemporal if the 
voltage and current variables depend both on time and space.

\begin{figure}[!t]
\centering
\includegraphics[width=3.5in]{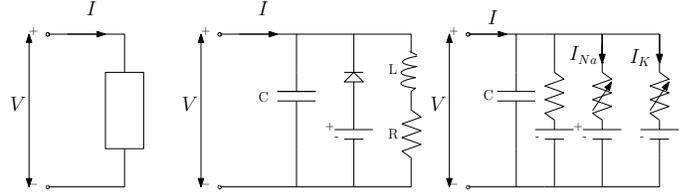}
\caption{Spking models have a circuit representation. (i) One-port circuit (ii) Nagumo circuit \cite{Nagumo1962} (iii) Hodgkin-Huxley circuit \cite{Hodgkin1952}}
\label{fig:spiking_circuits}
\end{figure}

The negative resistance circuit shown in Figure  \ref{fig:spiking_circuits} is a basic electronic circuit capable of switching, oscillating, 
and spiking. The circuit is composed of the classical elements of passive linear circuits except for the tunnel diode, modelled as
a negative resistance. A variant of Van der Pol circuit \cite{vanderpol1926}, it  was proposed by Nagumo \cite{Nagumo1962} as an elementary
model of spiking circuit.   The mixed nature of the negative resistance acknowledges the mixed nature of spiking. 
A control theory of spiking systems can be formulated in the modelling language of negative resistance circuits \cite{Sepulchre2018}. 
Spiking circuits are port interconnections of elements that include restricted ranges of negative conductance
in addition to the classical monotone elements of circuit theory. The {\it mixed} dissipativity properties of negative resistance
circuits is essential to the physical modeling of spiking circuits.

\subsection{Conductance-based modeling}

Since the pioneering work of Hodgkin and Huxley\cite{Hodgkin1952}, conductance-based modeling has been
the preferred mathematical language of biophysical neuronal networks. 

The conductance-based model of a neuron is a circuit composed of one capacitance (which models
the passive neuronal membrane), in parallel with possibly many current sources. Each current
source is modeled as the series-interconnection of a battery and a voltage-dependent conductance.
The resulting model of a current source is 
$$ I = G (V-V_0)$$
where $V_0$ is a constant (the battery (or Nernst) potential) and $G$ is the voltage-dependent conductance. The conductance
 $G$ is called internal if it only depends on the internal voltage $V$, whereas it is external if it
 depends on the voltage of other neurons. Internal conductances model the ion channels that regulate
 the flow of ions across the membrane, for instance the sodium  current $I_{Na}$ and potassium current $I_K$
 of Hodgkin-Huxley model. External conductances model the synaptic currents that depend on a pre-synaptic voltage.
 In addition to the source currents above, neurons can be interconnected by resistive wires, which model
 gap junctions.

\begin{figure}[h]
\centering
\includegraphics[width=3.5in]{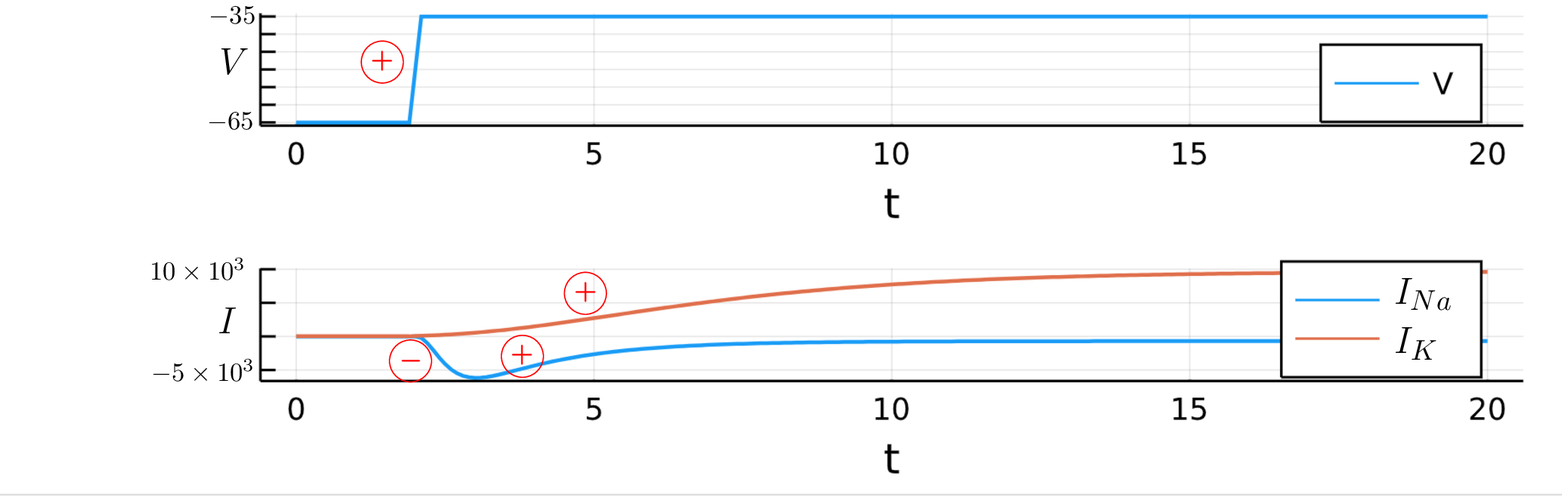}
\caption{The experimental step responses of  the two ionic currents of Hodgkin-Huxley model : the potassium current response is monotone,
whereas the sodium current response  is a difference of two monotone responses.}
\label{fig_HH monotonicity}
\end{figure}

The voltage dependence of ionic and synaptic conductances is nonlinear and dynamical. It models the mean-field
of populations of ion channels at the molecular scale and the empirical input-output relation is determined  from voltage-clamp
experiments. The key modeling insight from Hodgkin and Huxley came from separating the distinct contributions
of sodium and potassium channels and observing the monotonicity properties of the two currents for 
voltage steps of different amplitudes: the step response of the potassium current is always monotone,
whereas the step response of the sodium current is mixed-monotone, with a restricted voltage range
of {\it negative} conductance. This qualitative difference is illustrated in Figure \ref{fig_HH monotonicity}.

Those experiments justify the assumption that the potassium current source defines a monotone operator
on the space of voltage signals, whereas the sodium current source defines a {\it mixed}-monotone operator. Those assumptions extend to all current sources recorded in neurophysiology.
The mixed nature of conductances owes to distinct kinetics and voltage ranges for the 
activation and inactivation of the channel molecular gating. Conductances are monotone when
only the activation or inactivation process is modelled.  Likewise, 
the negative conductance of the sodium current $I_{Na}$ is key to the spiking behavior of Hodgkin-Huxley circuit \cite{Hodgkin1952}.

FitzHugh \cite{Fitzhugh1961} and Nagumo \cite{Nagumo1962} were first in recognizing the close analogy
between the circuit of Hodgkin and Huxley and the circuit of Van der Pol. The equations of FitzHugh-Nagumo circuit
are 
\ee\label{EQ: FNmodel}
\begin{array}{rcl}
C \dot V &  =  & k V - \frac{V^3}{3} -   I_L + I_{ext} \\
L \dot  I_L &  = &  - I_L + R V
\end{array}
\eee
which corresponds to the port interconnection of a linear capacitor, a leaky inductor (RL branch), an external current source,
and a negative resistance device.  The capacitive and inductive branches of the circuit are monotone. The negative resistance device
 $I= V^3 - k V$ is instead the difference of two monotone resistors. 

The negative resistance element of FitzHugh-Nagumo circuit is a simplification of the sodium current source of Hodgkin-Huxley model,
while the RL branch is a simplification of the potassium current source. There are however complications with the classical state-space
representations of conductance-based models. For instance, the state-space model of the potassium current in Hodgkin-Huxley model  defines
an input-output operator that is 
{\it not} monotone, even though it was fitted from monotone input-output experimental data. These issues are technical
and  beyond the scope of the present paper, but the interested reader is referred to \cite{vanwaarde2021} for a discussion. 

\subsection{A caveat about the meaning of negative conductance}

There is a persistent confusion in the literature about the meaning of negative conductance. This is because the same
terminology is used to refer to the  conductance $G$ of the model $I=G(V-V_0)$ and the {\it differential} conductance
$\delta I = g \, \delta V $. The latter refers to a variational (infinitesimal) quantity. The conductance $G$ is always positive, 
meaning that the sign of the current is positive (or {\it outward} in neurophysiology) for $V \ge V_0$ and negative (or {\it inward}
in neurophysiology) for $V \le V_0$. In contrast the differential conductance $g$ can be either positive or negative. In biophysical
models of neurons, this quantity is  dynamic and voltage dependent, hence it can be negative in a restricted   temporal and voltage
range, and positive elsewhere. The operator associated to a current is monotone if the differential conductance $g$ is always positive.

\section{Mixed feedback}
\label{sec:mixedfeedback}
\subsection{The mixed feedback amplifier}

The mixed feedback amplifier is an old electronic device that features the cover of the classical nonlinear
circuit textbook \cite{chua1987}. It provides a simple way to build negative resistance devices
from operational amplifiers (or, in modern times, from transistors), by wiring the output of the
amplifier both to its negative port (negative feedback) and to its positive port (positive feedback).
The device has an intimate connection to the history of control theory, because the stability of
amplifiers under feedback was a central drive to the early developments of the field.

{\it Positive} feedback amplifiers go back to the early days of electrical engineering. They provided
the early analog realizations of circuits that could switch and oscillate. The interest in {\it negative} feedback amplification
appeared in the 30's, after the groundbreaking discovery that high-gain negative feedback could considerably reduce
the uncertainty of the open-loop amplifier \cite{black34}. Understanding when and how negative feedback could also 
destabilize a system provided a key drive for the development of control theory. 

\begin{figure}[h]
\centering
\includegraphics[width=3.5in]{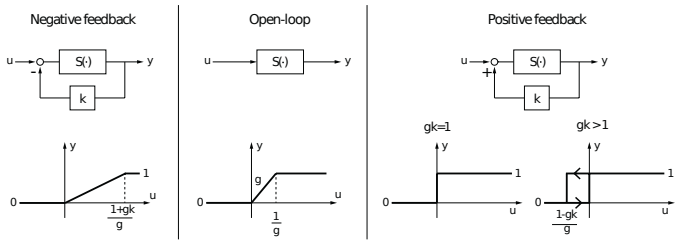}
\caption{Mixed feedback modulates the input-output behavior of an amplifier between a continuously modulated
system  (left) and a binary automaton (right). The behavior exhibits ultra-sensitivity at the boundary between
the analog and digital characteristics}
\label{fig:mixed-feedback}
\end{figure}

The  conceptual significance of mixed feedback for spiking control systems is  illustrated in Figure
\ref{fig:mixed-feedback}, which uses the simplest model of an operational amplifier: a saturating static model
\ee
\label{sat}
y=\sat_g(u) =  \left \{ \begin{array}{c} 0, \; u \le 0 \\ g \; v, \;  0 \le u \le \frac{1}{g} \\ 1, \; u \ge  \frac{1}{g} \end{array} \right .
\eee
in the feedback configuation shown in Figure
\ref{fig:mixed-feedback}. The parameter $g$ is called the open-loop gain and the parameter $k$ is called the feedback gain.  The open-loop model is $y=\sat_g(u)$. It is a linear model with gain $\frac{\delta y}{\delta u}=g$ over the restricted range $u \in [0,\frac{1}{g}]$. Away from this linear range, the  process saturates and the output is insensitive to input variations. The feedback model is defined by the implicit relationship $y=\sat_g(\pm ky+u)$. In a  negative feedback configuration, the relation rewrites as $y=\sat_{g'}(u)$ with $g'=\frac{g}{1+gk}$. The feedback relation is similar to the open-loop model, but with a lower gain over a broader linear range. Instead, in a positive feedback configuration, the linear range decreases and the gain increases. The linear range shrinks to zero for the critical value $k=\frac{1}{g}$, making the behavior ultra-sensitive. For larger positive feedback gains $k>\frac{1}{g}$, the closed-loop model is multivalued over the range $u \in [\frac{1}{g}-k,0]$. The output has then a binary readout for every value of the input. A continuous variation of the input signal leads to discontinuous jumps of the output signal between $0$ and $1$, with hysteresis. Positive feedback has converted an open-loop memoryless process into a closed-loop hysteretic relay, that is, a circuit realization of a digital bit. 

The mixed feedback amplifier captures the essence of a physical spiking system. The balance of mixed feedback
gains tunes the mixed-feedack device into anything between a continuously
regulated physical system and a digital binary automaton. The mixed-feedback amplifier is a design
paradigm for behaviors that exhibit the mixed nature of spiking signals. It cannot be categorized as either a physical control system or as a computational automaton,
because it is both. It combines the adaption of a continuous-time physical system and the reliability of a digital automaton.

The static model  also captures that an adequate balance of positive and negative feedback
results in a discontinuous or ultrasensitive input-output characteristic. This  boundary
between the analog and the digital world is the location of thresholds, a central feature
of spiking control systems.

\subsection{Mixed feedback systems}

\begin{figure}[h]
\centering
\includegraphics[width=3.5in]{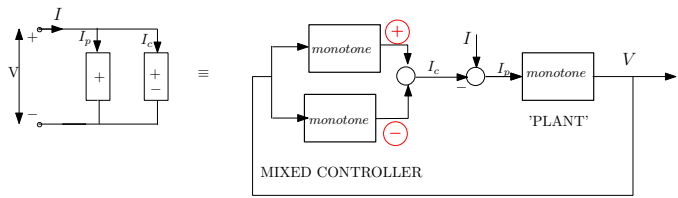}
\caption{The mixed-conductance circuit of a spiking control system and the corresponding mixed-feedback block-diagram  .}
\label{fig:mixed-feedback-system}
\end{figure}

The block diagram in Figure \ref{fig:mixed-feedback-system} provides a system theoretic generalization of the mixed feedback amplifer.
Its architecture is the one of a classical control system: the output of a physical plant, determined by a port relationship (voltage and
current in the case of an electrical circuit),  is fed back into a controller, that alters the input of the plant to shape the closed-loop response.
The feedback system defines a port interconnection between the plant and the controller, determined by Kirchhoff's law $I=I_c+I_p$.
This port interconnection shapes the new  ("closed-loop") port relationship between current and voltage.

What distinguishes Figure  \ref{fig:mixed-feedback-system}  from a classical control system is the {\it mixed} nature of the controller. The controller is 
not monotone. Instead, it is the difference of two monotone operators. The monotonicity of each block can be understood as
a sign preserving property: a positive increment at the input implies a positive increment at the output. As a result, the block diagram generalizes 
the concept of mixed-feedback:  the mixed controller generates two parallel feedback loops of opposite sign.

The negative feedback loop is the feedback loop of conventional control, where both the plant and the controller are assumed to be monotone.
The function of the negative feedback loop is a generalization of the negative feedback amplifier: it reduces the sensitivity of the output both
to external disturbances and to plant uncertainty. The positive feedback loop is the feedback loop that turns the physical plant into an automaton.
The function of the positive feedback loop is a generalization of the positive feedback amplifier: it makes the output digital, that is, one out
of a finite set of discrete states. 

The reader will notice that a mixed feedback system does not necessarily define an operator. However the inverse system is always well defined as a difference of monotone operators.
Studying the input-output trajectories as solutions of the inverse operator is a key insight of the present paper.

\subsection{Spiking: {\it first} positive {\it then} negative feedback}

The mixed feedback nature of the control system in Figure  \ref{fig:mixed-feedback-system} acknowledges the mixed nature of spiking
control system.  But the {\it static} model of mixed feedback amplification fails to  model the {\it dynamical} nature of spiking: spiking is a temporal {\it event}, 
that results from a {\it transient} switch,  {\it localized} in time, amplitude, and space. To transform the static element in a spiking device,
one must model the  dynamical {\it hierarchy} between positive and negative feedback. The destabilizing positive feedback must come {\it first}, {\it then} the repolarizing
negative feedback may kick in. This hierarchy results from a positive feedback gain localized in a high-frequency temporal range and a tiny amplitude range,
relative to the negative feedback gain that dominates the positive feedback gain away from the localized range. The localized nature of an event results from
this hierarchy:  the fast positive feedback allows for the switch,
whereas the slow negative feedback makes it transient. The fast positive is necessary for the digital reliability of the spike time, whereas
the slow negative feedback is necessary for the analog regulation of the event. 

Each negative conductance element in the controller circuit of a spiking control system controls one such excitability  mechanism. The signature
of this mechanism is the existence of a threshold. The threshold results from a point of zero conductance in the total loop gain. Each threshold
 results from a localized temporal and amplitude window over which the total negative conductance of the controller outweighs the 
total positive conductance. The balance of positive and negative feedback is a generic mechanism to create points of zero conductance. This
mechanism is robust to the uncertainty of system components, provided that the maximal conductances of the elementary current sources
can be adapted. This is the essential role of neuromodulation.

\begin{figure}[h]
\centering
\includegraphics[width=2.5in]{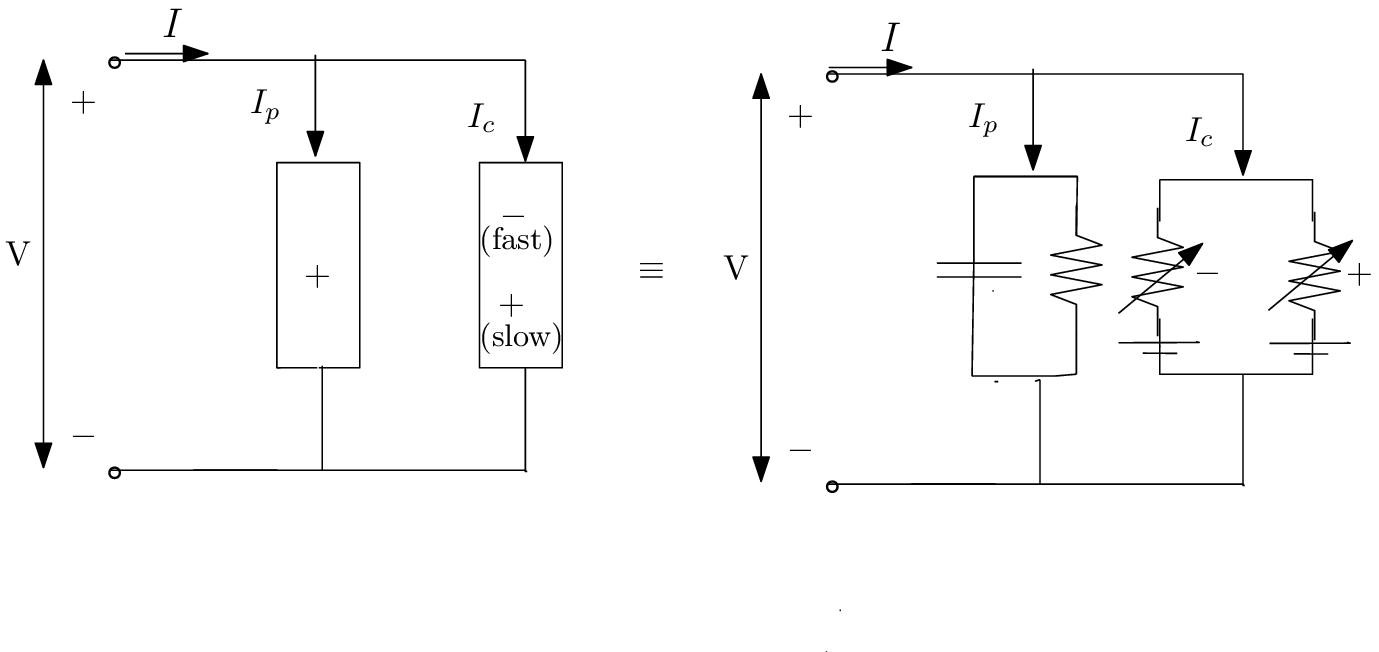}
\caption{The circuit representation of an elementary spiking circuit: the monotone "plant" is a passive RC circuit, and the mixed "controller" is
the parallel connection of a negative conductance element and a positive conductance element. The negative conductance  only dominates the positive
conductance over a narrow amplitude and temporal ranges, thereby creating an excitability threshold.}
\label{fig:mixed-circuit}
\end{figure}

Both FitzHugh-Nagumo and Hodgkin-Huxley circuits provide physical realizations of the mixed-feedback block diagram. In both instances, the
plant is a RC circuit modeled as a linear time-invariant passive plant. This first-order lag model is the prototype model of a leaky integrator
with an exponentially fading memory. 

In  FitzHugh-Nagumo circuit, the controller is the parallel interconnection of a negative resistor
with a RL filter. The restricted region of negative conductance of the nonlinear resistor provides  the negative conductance element of 
the controller. The positive parameter $k$ in Equation (\ref{EQ: FNmodel}) controls the threshold. The circuit is a spiking control circuit in the relaxation regime, when the time-constant of the controller (inductive branch) is much
larger than the time constant of the plant (capacitive branch).

In Hodgkin-Huxley model, the controller is the parallel interconnection of the sodium current  and the potassium current. The activation
of sodium channels provides the negative conductance of the circuit. The spiking nature of Hodgkin-Huxley model results from 
the {\it fast} activation of sodium channels relative to the {\it slow}  activation of potassium channels and {\it slow} inactivation of sodium channels.
Those qualitative distinctions are rather clear from the step response shown in Figure \ref{fig_HH monotonicity}.

The dynamics of an elementary spiking circuit are often further approximated by an integrate-and-fire model. The modelling framework
in this paper insists on integrate-and-fire models that have the physical realization of a port nonlinear circuit, but applies to
circuit realizations that include diodes or digital transistors. Monotone operators include the mathematical description of such
discontinuous behaviors, see e.g. \cite{Camlibel2016,Brogliato2020}. The relationship between conductance-based models and 
physical integrate-and-fire models is further discussed in \cite{VanPottelbergh2021}.

The block diagram in Figure \ref{fig:mixed-feedback-system}  is a general representation of spiking control systems, both
valid for single neurons ($I$ and $V$ scalar variables) and for neuronal networks ($I$ and $V$ vector variables). The conductance-based model
of an arbitrary neuronal network  obeys the natural decomposition between a passive plant modeled by a RC network
of $N$ capacitances interconnected by resistive wires, and a mixed controller gathering all the active current sources.
In biophysical terms, the plant represent the passive membranes interconnected by gap junctions, whereas the mixed
controller includes all voltage-gated ionic and synaptic currents. 

\subsection{Interconnections}
\label{sec:interconnections}

The mixed motif circuit in Figure \ref{fig:mixed-circuit} is a basic circuit element of
spiking control systems. More complex spiking systems result from port
interconnections of this basic motif.  Neurophysiology suggests a highly
modular and hierarchical architecture of such interconnections in the animal
world.  This is not so different from man-made control systems. An elementary 
physical system such as an electrical motor is controlled with a two-parameter lead-lag controller
and provides a basic motif for a control systems. A complex control system such as a power plant
proceeds from  a hierarchical interconnection of elementary control systems.
The localization of the individual control systems in specific temporal and amplitude
windows is key to the hierarchical architecture of the control system.

The architecture of spiking control systems in the animal world is best documented for
small circuits (central pattern generators) that control  rhythmic functions such as respiration, chewing, or locomotion.
One of the most extensively studied spiking control systems in the animal world
is the stomatogastric ganglion (STG) of the lobster \cite{STG:scholarpedia}.
The neuromodulatory control system of this network of about thirty interconnected neurons
is remarkable by its complexity and its hierarchical organization \cite{Marder2018}. 

The core mechanism of rhythm generation in central pattern generators is a network
of rebound bursting cells interconnected by inhibitory synapses. A burst in a presynaptic neuron
hyperpolarizes the membrane of the postsynaptic neuron, which in turns exhibits
a rebound burst when the presynaptic neuron releases its inhibition.

\begin{figure}[h]
\centering
\includegraphics[width=2.5in]{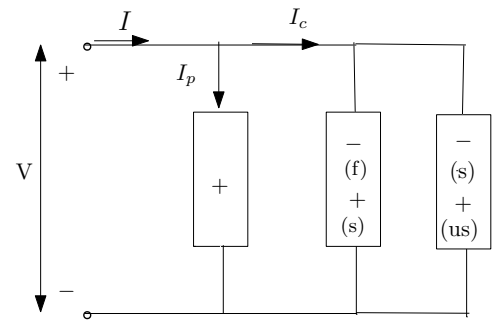}
\caption{The circuit of a bursting neuron repeats twice the mixed-conductance controller of a spiking neuron. The two 
negative conductances create two distinct localized amplitude and temporal ranges where the total negative conductance
of the circuit dominates the total positive conductance. The resulting circuit has two rather than one excitability thresholds,
which distinctly control the fast (intraburst) and slow (interburst) oscillations of bursting. }
\label{fig:bursting}
\end{figure}

A bursting one-port circuit is obtained by the port interconnection of two spiking one-port circuits \cite{Franci2017,sepulchre2019}.
The hierarchy between the two port circuits is both in amplitude and in time: bursting results from interconnecting
a {\it fast} spiker with {\it high} threshold and a {\it slower} spiker with {\it lower} threshold. The bursting circuit has the block
diagram representation of Figure \ref{fig:bursting}, with a controller made of four parallel conductances:
a fast mixed conductance, a slow positive conductance, a slow mixed conductance, and an ultra-slow positive conductance.
The hierarchy of the corresponding currents is well documented in neurophysiology, see
\cite{Franci2017} for details.

\begin{figure}[h]
\centering
\includegraphics[width=2.5in]{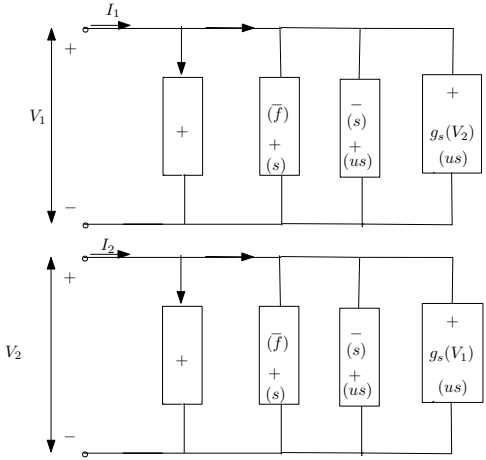}
\caption{The circuit of a half-center oscillator interconnects two bursting neurons through port
interconnection of each neuron to a synaptic conductance that depends on the other neuron. The
ultraslow positive conductance of each synapse control the antiphase rebound bursts of the two neurons.  }
\label{fig:HCO}
\end{figure}

Inhibitory synaptic coupling between two bursting cells is the simplest architecture of a rhythmic circuit. This
circuit has been studied under the name "Half Center Oscillator (HCO)" for more than a century \cite{brown1914}.
The two rebound bursters do not burst in isolation, but can sustain an antiphase rhythm through their mutual
inhibition. The circuit is realized by the port interconnection of a synapse to each bursting neuron. The resulting
circuit is shown in Figure \ref{fig:HCO}. It defines a mixed-monotone relation between two input currents and two output voltages. 

A central pattern generator such as the stomatogastric ganglion controls the interaction between
a fast and a slow rhythm. Each rhythm can be modeled as a simple HCO, and the two
HCOs interact through a central hub node. The versatile control of this mixed-feedback system is
studied in \cite{Drion2018b}.  It will be illustrated in Section \ref{sec:neuromodulation}.

\section{Spiking system analysis}
\label{sec:analysis}
\subsection{Monotonicity and feedback}

Monotonicity, or incremental passivity,  is a key system property for feedback system analysis. 
Its importance stems from two fundamental properties : the sum of
two monotone operators is monotone, and the inverse of a monotone
operator is monotone. 

Consider the feedback system in Figure and assume that both the plant $P$ and the
controller $C$ define monotone operators. The relationship between a current signal $I$
and a voltage signal $V$ solutions of the feedback system must satisfy the relation
$$  I = P^{-1}(V) + C(V),$$
which defines a monotone operator from $V$ to $I$ if $P$ and $C$ are monotone.
The inverse of that operator is a monotone operator from $I$ to $V$, and it
characterizes all the solutions of the feedback system. This result
 is called the incremental passivity theorem in control theory \cite{Desoer1975}. It is
a pillar of feedback system analysis, showing that the negative feedback interconnection
of monotone systems defines a closed-loop monotone system. The relationship is
also the port interconnection defined by the parallel interconnection of two electrical
circuits, that is, the physical interconnection of a physical plant and a physical controller. 

Beyond its significance for physical feedback interconnections, monotonicity also
paves the way to an  {\it algorithmic} treatment of analysis and synthesis questions.  
This is best illustrated by the most basic question of computing the input-output solutions
of the feedback system. Given a current trajectory $I^*(\cdot)$, determining the corresponding
closed-lop voltage trajectory $V^*(\cdot)$  amounts to solve 
$$ 0 \in P^{-1}(V) + C(V) - I^* $$
which is  the problem of determining a zero of a monotone operator in the signal space of voltages. This question 
the core algorithmic question of convex optimization. It can be solved
efficiently with first-order iterative algorithms that scale up to large-scale and/or non-smooth 
problems. The reader is referred to \cite{chaffey2021} for more details. Monotonicity
has been exploited  in the algorithmic analysis of physical nonlinear systems,
see for instance \cite{Brogliato2020} and \cite{Camlibel2016}.

The control theory of monotone feedback systems is best developed for LTI systems,
in which case monotonicity is equivalent to passivity. The Kalman-Yakubovich lemma establishes
a bridge between the positivity of the operator in the frequency-domain and
the solution of a Linear Matrix Inequality for a state-space realization of the operator. 
This bridge is key to the solution of most analysis and design questions of linear control theory
via convex optimization \cite{boyd1994}. The theory of maximal monotone operators paves the
way for a generalization of those algorithmic results to nonlinear systems.

\subsection{Solutions of a mixed feedback system}

In a spiking control system, the controller is not monotone. Instead, it is the difference
of two monotone operators  $C_1$ and $C_2$, that is, $C=C_1-C_2$.
Mimicking the development in the previous section, finding the input-output solutions of a mixed feedback system 
amounts to solve
\ee
\label{mixed-monotone}
0 \in \underbrace{P^{-1}(V) + C_1(V) - I^*}_{\text{monotone}} - \underbrace{C_2(V)}_{\text{monotone}}
\eee
which is the problem of finding a zero of a {\it difference} of two monotone operators. This is the core algorithmic
question of "difference of convex" (DC) programming, which
aims at minimizing a  nonconvex cost function expressed as the difference of two convex functions.  While DC programming 
is too general to be tractable without further assumptions, "disciplined" DC programming has proven useful at
 leveraging tools of convex optimization to specific structured non-convex optimization
problems \cite{Shen2016}. The  non-monotone zero finding problem (\ref{mixed-monotone}) 
is solved iteratively via a sequence of monotone zero finding problems of the type 
\ee
\label{mixed-monotone}
0 \in P^{-1}(V_{i+1}) + C_1(V_{i+1}) - I^* - C_2(V_i), \; i = 0,1, \dots
\eee
The recent paper  \cite{das2021} illustrates the potential of this iterative algorithm on the elementary example of FitzHugh-Nagumo
model where the plant $P$ is a LTI passive system, the monotone controller $C_1$ is the sum of a passive LTI
system and a cubic static nonlinearity, and the controller $C_2(V)=kV$ is linear and static. 

The convergence analysis of such an iterative algorithm is only local, but it is a disciplined departure from the global
convergence analysis of a monotone operator: the convergence analysis is global for $k=0$, and
the global solution of the monotone operator can be deformed by continuation for $k \neq 0$.

The structure of spiking control systems is particularly suited to leveraging the analysis tools
of monotone feedback systems to mixed  feedback systems. This is because
the decomposition of the controller $C$ as the difference of two monotone operators matches
the modeling distinction between positive and negative conductances. The controller $C_2$ 
only includes the negative conductances of the spiking circuit. By design, those are
few and localized: each negative conductance controls one threshold. Biological
control systems suggest an architecture composed of a realm of positive conductance elements
for {\it regulation} and a few negative conductance elements, each of which controls the threshold
of a discrete event. Such an architecture calls for an analysis framework in which the departure
from monotonicity is structured and has a precise physical interpretation.

The relationship between monotonicity and convexity is useful to appreciate the conceptual difference between
 a monotone feedback and a mixed feedback system. Suppose that each monotone operator in the block-diagram of Figure \ref{fig:mixed-feedback}
is the gradient of a convex function. If the feedback system is monotone, then each input-output trajectory
is the minimizer of a convex function.  Instead, the addition of a negative conductance element in the controller
opens the possibility of multiple output trajectories for a given input trajectory: the
addition of a localized concave function to a convex function creates a double well potential. 

Mixed feedback systems offer a departure of fading memory control systems to control systems 
equipped with localized threshold and  memories. Each negative conductance element of the 
feedback system can be imagined as  a localized bump in an otherwise convex landscape. This image suggests
the algorithmic potential of  disciplined DC programming \cite{Shen2016}  in the analysis of spiking control systems.

\subsection{Output feedback monotonicity}
\label{sec:adaptive}

An elementary but key property of the mixed feedback system in Figure \ref{fig:mixed-feedback-system}
is output feedback monotonicity: the output feedback transformation $I=-K V + I_{aux}$ renders the system
monotone, or incrementally passive,  from $I_{aux}$ to $V$ provided that the gain $K>0$ exceeds the maximal gain of the negative conductance
of the mixed controller. Physically, this means that any spiking control system is turned into a monotone system
by attaching a resistive port to the circuit. The transformed circuit has infinite gain margin, that is, cannot be destabilised
by (negative) output feedback, and has a contractive inverse. In classical control theory, such systems are the simplest to control.
They lead to simple designs for synchronization, observer design, adaptive control, and adaptive parameter estimation.

It is of considerable interest that systems that exhibit switches and oscillations can be modelled as simple control systems. 
As an illustration, the observer (or synchronization) problem has an elementary solution: the observer can be designed
 as a mere copy of the spiking control system, with 
input $I$ and output $\hat V$. The error $\Delta V = V -\hat V$ is fed
back to the input of observer, and the observer trajectories contract to the system trajectories provided that the
feedback gain is sufficient. 

Output feedback monotonicity is further explored in the recent paper \cite{burghi2021a}
to solve the {\it adaptive} observer problem: the unknown parameters of the observer are the maximal
conductances of the ionic and synaptic currents. The resulting spiking control system is linear in the
unknown parameters, and the solution of the adaptive observer is a classical least-square recursive
algorithm of adaptive control. This adaptive observer has global properties and can track in real 
time the neuromodulation of a spiking control system. It paves the way to the design of adaptive internal
model controllers. This property illustrates that spiking control systems inherit the adaptation functionalities
of continuous-time physical systems.

\section{Designing spiking control circuits}
\label{sec:design}
It is one thing to develop an analysis framework for the spiking control systems
encountered in the animal world, but what is the significance of  designing
 {\it artificial} spiking control systems ?
At first glance, such a question belongs to the pre-digital age. It was a central quest
of Cybernetics \cite{ashby1960}. The homeostat \cite{wiki:homeostat} of Ross Ashby is an example of analog
machine that switches between discrete states and at the same time
continuously adapts to its environment.  

According to Wikipedia \cite{wiki:homeostat}, the homeostat did not work very reliably. 
It suffered the fundamental limitation of analog systems, which stems from their sensitivity to the  uncertainty
of the system components. Resistors are sensitive to temperature, and no two
amplifiers have the same analog range. Analog systems are inherently sensitive
to the uncertainty and variability of their components. In neuromorphic engineering,
this challenge is known as the transistor mismatch \cite{Poon2011}. 
Analog designs seem intrinsically unreliable.

The digital age solved the unreliability of analog computation by building 
devices that can only switch between the discrete states of an automaton.
Digital technology is reliable, at the price of quantization. Time is quantized
according to a clock, and amplitude is quantized according to the low or
high voltage of a transistor. 

To some extent, the neuromorphic dream is a renaissance of the cybernetics dream: 
it aims at designing physical machines that combine
the adaptation of analog systems and the reliability of digital
automata. This section briefly discusses how those two objectives can be mixed
and why such systems might be needed in the post-digital technology.

\subsection{Mixing analog variability and digital reliability}

Reliability and adaptation are mutually exclusive both in the worlds
of automata and physical systems: digital automata are reliable {\it because} they do
not change continuously, and physical systems are  unreliable {\it because}
they are continuously adaptive. 

Thanks to their mixed nature, spiking control systems combine adaption and reliability.
They inherit the adaptation of analog systems in their subthreshold
regime, but the timing of their discrete events inherits the reliability of automata.

\begin{figure}[h]
\centering
\includegraphics[width=3in]{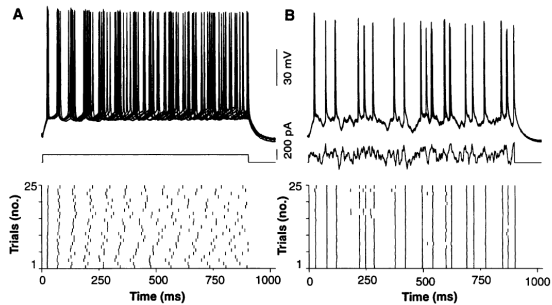}
\caption{The reliability experiment reproduced from \cite{mainen1995}. }
\label{fig:reliability}
\end{figure}

This property is best illustrated by a famous experiment, first conducted
in Aplysia neurons by Bryant and Segundo in 1976 \cite{Bryant1976}, and 
then beautifully reproduced in neocortical neurons by Mainen and Sejnowski in 1996 \cite{mainen1995},
see Figure \ref{fig:reliability}.

The same input-output experiment is repeated 25 times on the same neuron, in the
same experimental conditions. In response to a step change of current, the neuron
switches from a resting state to a spiking state. However, the spiking rhythm is
variable from experiment to experiment. In sharp contrast, a same fluctuating input
current repeated ten times results in a reliable sequence of spike timings. 
 
 The first part of the experiment exhibits the variability of an analog circuit. The
spiking attractor is unreliable because it is adaptive. It is sensitive to small
 changes in the environment from experiment to experiment. The second part
 of the experiment exhibits the reliability of discrete events associated to the threshold
 property of the spiking neuron. The same fluctuating input causes the same sequence
 of supra-threshold responses, ensuring reliable signal transmission.

The reliability experiment illustrates why the design of a spiking control system differs
from the design of a conventional control system. Spiking control systems should be
designed to combine digital reliability and analog adaptation.  The digital reliability stems
from threshold properties, and each threshold is controlled by one negative conductance
element of the circuit. The analog adaptation stems from the sensitivity of the input-output
behavior to continuous parameter changes in all the  conductance elements of the feedback control system.

The organization of spiking control systems in the animal world suggests that the
variability and uncertainty of system components is a feature rather than 
a limitation of their analog nature. Animal nervous systems combine digital reliability and analog  adaptation.
The mixed architecture of spiking control systems
seems essential to this mixed performance and a key motivation for the design of
 control systems that can combine analog adaptation and digital reliability.
 
The reliability experiment was reproduced in silico in the recent work \cite{kirby2021}. We
observed the exact same phenomenon in the neuromorphic circuit implementation
of a half-center oscillator, reproducing the co-existence of reliability and adaptation
at  three distinct hierarchic scales:  single neuron spiking, single neuron bursting, 
and the rebound rhythm of the HCO.

\subsection{Event-based control}

In the light of the digital age, the invention of the computer made mixed-feedback systems obsolete.
For the past seventy years, digital technology has emulated with astonishing successes
the performance of analog systems with the reliability of automata. However, this performance requires 
 ever faster digital clocks, ever finer quantization, and comes at the price of an ever 
increasing energy cost.

The neuromorphic proposal of Carver Mead is to turn
digital computers  into event-based analog machines.
The temporal resolution of events is not restricted by a digital clock, and
the amplitude resolution of events is not restricted by bits. Instead, the temporal
and amplitude resolution of the events is {\it adaptive} 
through classical averaging and ensemble mechanisms.

\begin{figure}[!t]
\centering
\includegraphics[width=2.5in]{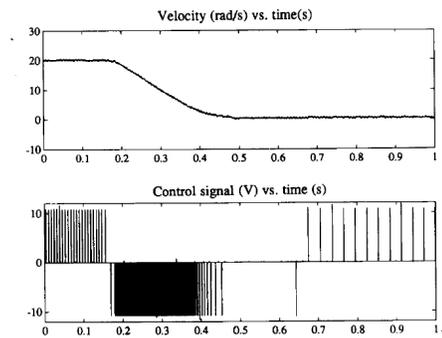}
\caption{Spiking control of an electrical motor. One same spiking control algorithm emulates
both the stepper motor needed at low speed and the PWM servo needed at high speed. 
The stepper motor functions as an event-based automaton, the PWM controller functions as a 
continuous regulatory system through averaging. (Reproduced from \cite{deweerth1991})}
\label{fig:DCmotor}
\end{figure}

Even at the elementary level of a one port circuit, the event-based nature
of a spiking control system sheds new light on simple control problems.
A compelling illustration is provided by the early encounter of Carver Mead and Karl Astrom. 
The article  \cite{deweerth1991} demonstrates the distinct properties of a spiking controller
in the most elementary  problem of regulating the speed of an electrical motor. 
 Figure \ref{fig:DCmotor} illustrates that the spiking control system functions
 as a classical pulse-width-modulated controller at nominal speed, 
 whereas it functions as a stepper-motor at very low speed, when each spike
 overcomes the dry friction and turns the motor by a small angle. With
 a  conventional controller, the motor would stop at the low reference speed
 because of friction. 
 
 In the terminology of neuroscience, the spiking controller transitions
 from a {\it rate} code to a {\it spike} code as the scale of the reference speed
 varies. This feature illustrates the remarkable adaptation of spiking control 
 systems across scales. In the digital age, the technology of PWM controllers
 and stepper motors obey different design principles.  
  
One decade after his encounter with Carver Mead, Karl Astrom developed the
concept of event-based control \cite{astrom2008}.  The proposed model of 
spiking control systems suggests that classical control theory can be leveraged
to  the physical design of event-based control systems

\subsection{Neuromodulation across scales}
\label{sec:neuromodulation}
Neuromodulation is a key component of the design architecture of spiking control systems.
In neurophysiology, it designates the realm of biochemical mechanisms that can modulate
or adapt the expression or gating of specific ion channels. To appreciate the significance of
neuromodulation in animal spiking control systems, the interested reader is referred
to the seminal work of Eve Marder (e.g. \cite{Marder2012,Marder2018} and
 references therein). In the context of this paper,
it is sufficient to think of neuromodulation as the possibility of modulating or adapting the
maximal conductance of any internal or external current of the controller.

Neuromodulation endows spiking control systems with all the capabilities of 
traditional adaptive control \cite{Astrom2008b}, as for instance illustrated by the online estimation scheme 
discussed in Section \ref{sec:adaptive}. However, they also exhibit distinctive adaptation
capabilities  that are specific to the modulation
of {\it negative} conductance parameters. Such capabilities have not been explored in the theory of adaptive
control. We focus on those in the rest of this section, and refer
the reader to  \cite{sepulchre2019,Drion2018,Drion2018b} for further details.

We have stressed in Section \ref{sec:mixedfeedback} that individual negative conductances control individual {\it thresholds},
which themselves control the spatial and temporal scales of {\it events}. For this reason,
the adaptation of negative conductances is a distinct adaptation mechanism that controls the 
modulates the {\it
scale} of thresholds, thereby enabling a unique control mechanism {\it across} scales.

\begin{figure}[!t]
\centering
\includegraphics[width=3.5in]{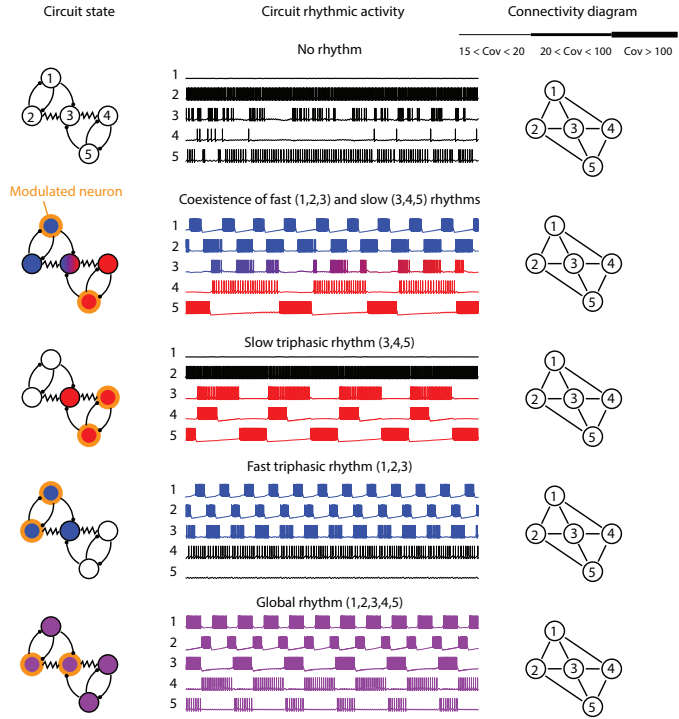}
\caption{Control across scales: the discrete states of a rhythmic {\it network}
are controlled via parameter modulation of {\it cellular} parameters.
Each neuron controls its participation in the network rhythm
via the modulation of a single parameter: the maximal conductance of its slow negative
conductance current  (Reproduced from \cite{Drion2018b})}.
\label{fig:STGtopology}
\end{figure}

We illustrate this general and far reaching principle on the simple central
pattern generator discussed in Section \ref{sec:interconnections}. We saw that rebound bursting 
neurons have both a fast (spiking) and a slow (bursting) thresholds. The slow threshold 
is a critical parameter for the rebound properties of the neuron, which in turn determine
the rhythmic state of a central pattern generator when interconnected in an inhibitory network.
This suggests that modulating the slow {\it negative} conductance of a rebound bursting neuron
plays a critical role in controlling the rhythm of a network. This function is illustrated
in Figure \ref{fig:STGtopology}, reproduced from \cite{Drion2018b}: five different rhythmic states
of a same network are obtained by modulating a single internal parameter of each neuron,
namely  the maximal conductance of  the slow negative conductance controlling the rebound neuronal 
property.

This modulation is an example of control across scales: internal parameters at the cellular
scale are effective in modulating the functional topology of a rhythmic network. In fact, it is
further shown in \cite{Drion2018b} that achieving a similar network modulation with other
parameters, such as the maximal conductances of the synaptic interconnections, is both
more challenging and  more fragile. This observation is not accidental. It is the result of 
decoupling the role of negative and positive conductances in a mixed control architecture,
and acknowledging that negative conductances control the discreteness of events
whereas positive conductances regulate their analog properties. The distinct role
of negative and positive conductances in the adaptive control of a spiking
control system is a distinctive property of mixed feedback systems. 
The article \cite{Dethier2015} provided
an earlier account of the key role of slow negative conductances in controlling 
the rhythm of an elementary half-center oscillator.

The elementary design principle illustrated in Figure \ref{fig:STGtopology} 
enables a general network control principle across scales,  as the internal parameter modulation
at the cellular scale efficiently controls the functional connectivity of the network.
A same physical connectivity can be turned into a combinatorial
number of  distinct functional connectivities  depending on which node is turned on 
and which node is turned off. This principle is not confined to central pattern generators. 
A more general illustration is provided in \cite{Drion2018} on a spatially extended
network of 200 neurons inspired by the neurophysiology of brain states. Brain states
are interpreted as the discrete states of a large spiking control network. Each brain state 
carries  a specific spatio-temporal signature. The modulation of the cellular negative conductances
orchestrates the transition between them.

The neuromodulation of cellular negative conductances provides a basic
mechanism for the nodal control of a network topology. It opens
entirely novel design possibilities for the design and control 
of  networked event-based systems. The cellular
mechanism has been demonstrated in silico \cite{ribar2019} and provides
a key design principle for large-scale spiking networks.
 
Neuromodulation is a key design principle to make spiking control systems adaptive. 
There is a lot that remains to be explored in the application of neuromodulation in  {\it mixed}
feedback systems by acknowledging the distinct role of positive and negative
conductance parameters in controlling both the discrete and continuous
features of spiking behaviors.

\section{Back to the future}

This article has illustrated that spiking control systems can be analyzed and designed
by revisiting and leveraging classical tools from control theory. 

Spiking systems are modelled as nonlinear electrical circuits. They can be interconnected
and have a modular architecture. The essence of a basic spiking circuit is
that it requires elements with both positive and negative conductance.

Monotonicity provides the mathematical abstraction of a nonlinear resistor and is a foundation
for the analysis. Spiking control systems have the classical block diagram representation
of a control system that mixes positive and negative feedback loops of monotone operators. 
Negative feedback of monotone operators characterizes classical control, in which case the feedback system
itself defines a monotone operator. In spiking control, the mixed feedback loop is studied via a
difference of monotone operators. Maximal monotonicity paves the way to an algorithmic
analysis of spiking feedback systems. The departure from monotonicity is structured and
disciplined, similar to the departure from convexity in difference-of-convex programming. 

The design of spiking control systems provides a methodology for the physical realization 
of event-based control systems.  Event-based technology has flourished in the recent years. 
Event-based cameras revolutionize the technology of dynamic vision \cite{gallego2020}, and event-based sensors will
revolutionize the technology of dynamic grasping and touching \cite{Osborn2016}. The fast
development of those new sensors and actuators calls for the development
of event-based control systems. 

There is a pressing need for a better theory of spiking systems across medicine and engineering.  Brain-machine interfaces will determine the future of neuroengineering and neuromedicine.  They will require control systems that can be interconnected to natural neural systems, acknowledging the spiking nature of neural signals rather than concatenating neural signals with analog-to-digital and digital-to-analog interfaces. Whether in neuromorphic engineering or in neuroengineering, the control engineer should benefit from a unified framework  to model, analyze, and design natural or artificial spiking control systems.  We have argued that such a framework should acknowledge a mixed feedback principle as the essence of design principles that can combine the continuous adaptation of analog systems with the discrete reliability of digital automata.


%



\ifCLASSOPTIONcompsoc
  \section*{Acknowledgments}
\else
  \section*{Acknowledgment}
\fi
This work has benefited from many collaborators over the years. In particular, the author wishes to acknowledge 
the help and insight from Fulvio Forni, Timothy O'Leary, Malcolm Smith, Guillaume Drion, Alessio Franci,
Luka Ribar, Thiago Burghi, Tomas Van Pottelberghe,  Ilario Cirillo, Tom Chaffey, and Tai Kirby. 
The research leading to these results has received funding from the European Research Council under the
Advanced ERC Grant Agreement Switchlet n.670645.

\ifCLASSOPTIONcaptionsoff
  \newpage
\fi



%
\bibliographystyle{IEEEtran}
\bibliography{IEEEProceedings}

%
\begin{IEEEbiography}[{\includegraphics[width=1in,height=1.25in,clip,keepaspectratio]
{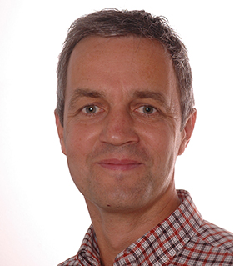}}]
{Rodolphe Sepulchre} 
(M96,SM08,F10) received the engineering degree and the Ph.D. degree from the
Universit\'e Catholique de Louvain in 1990 and in 1994, respectively.  He is
Professor of Engineering at the University of Cambridge since 2013.  His  research
interests are in nonlinear control and optimization, and more recently
neuromorphic control.  He co-authored the monographs ``Constructive Nonlinear
Control'' (Springer-Verlag, 1997) and ``Optimization on Matrix Manifolds''
(Princeton University Press, 2008). He is Editor-in-Chief of IEEE Control Systems.
He is a recipient of  the IEEE CSS Antonio Ruberti Young Researcher Prize (2008)
and of the IEEE CSS George S. Axelby Outstanding Paper Award (2020). He is a
fellow of IEEE, IFAC,  and SIAM. He has been IEEE CSS distinguished lecturer 
between 2010 and 2015. In 2013, he was elected at the Royal Academy of Belgium.
\end{IEEEbiography}




\end{document}